\documentclass[11pt,twoside]{article}

\usepackage{asp2004}
\usepackage{epsf}
\usepackage{psfig}
\usepackage{lscape}

\def\be{\begin{equation}}
\def\ee{\end{equation}}
\def\bea{\begin{eqnarray}}
\def\eea{\end{eqnarray}}

\def\solmas{{M$_\odot$}}
\def\simless{\mathbin{\lower 3pt\hbox
   {$\rlap{\raise 5pt\hbox{$\char'074$}}\mathchar"7218$}}}   
\def\simgreat{\mathbin{\lower 3pt\hbox
   {$\rlap{\raise 5pt\hbox{$\char'076$}}\mathchar"7218$}}}   
\def\etal{{\rm et al.}}

\def\solmas{{M$_\odot$}}

\def\etal{\rm  et al.}

\begin{document}
\title{The Formation of Massive Binary Stars}   
\author{Ian A. Bonnell}   
\affil{School of Physics and Astronomy, University of St Andrews, St Andrews, Scotland, KY16 9SS, UK}    

\begin{abstract} 
The formation of massive stars in close binary systems is complicated due to their high
radiation pressure, the crowded environment and the expected minimum separation for fragmentation being many times greater than the orbital separation. I discuss how massive star
formation can be understood as being due to competitive accretion in stellar clusters. 
Massive binary systems are then formed due to accretion onto wider low-mass
systems. The combination of accretion and dynamical interactions with
other stars in the cluster then result in producing a close binary system.  
Tidal and 3-body captures in dense stellar cores can also
play a role in forming massive binary systems while
stellar mergers
of such close binary systems due to interactions may play an important role in overcoming the radiation pressure.

\end{abstract}


\section{Introduction: The Problem}   

Although massive stars are commonly found in close binary systems, their
formation mechanisms are poorly understood.   
There are a number of difficulties that need to be overcome in order to explain
their formation mechanism. These include the intrinsic difficulty of forming
massive stars, and the intrinsic difficulty of forming close binary systems, even in
amongst low-mass stars.

\subsection{ Formation of Massive Stars}   

There are several potential difficulties in forming high-mass stars. Firstly, the
timescale of less than $10^6$ years to assemble 10 to more than 100 \solmas\  implies
large accretion rates. Secondly, their crowded location in the centre of young
stellar clusters implies that there was not enough room to fit in the precursor fragment.
In order to get the Jeans radius to be smaller than the interstellar separation, the 
resultant Jeans mass is also correspondingly small (Bonnell, Bate \& Zinnecker 1998).
Lastly, and most importantly, the radiation pressure from a high-mass star is sufficient
to reverse the infall of gas that contains typical dust properties (Wolfire \& Casinelli 1987; Beech \& Mitalas 1994).
There are a number of ways this last problem can be circumvented. The radiation
pressure can be overwhelmed by ultra-high accretion rates (McKee \& Tan~2003)
although how this in practice can occur is unclear. More promising is that accretion
occurs preferentially through a disc, combined with a rapidly rotating star that
emits most of its radiation towards the poles (Yorke \& Sonnhalter 2002). A third
(and fairly exotic) solution is that massive stars form due to stellar mergers in the
ultradense core of a cluster (Bonnell, Bate \& Zinnecker 1998).

\subsection{Formation of Close Binary Stars}   

Forming close binary stars systems is difficult even amongst lower-mass stars, The reason for this
is that if the components form through fragmentation (eg, Boss 1986; Bonnell 1999), then the Jeans radius at the
point of fragmentation must be smaller than their separation,
\be
R_{\rm sep} \simgreat  2 R_J \propto T^{1/2} \rho^{-1/2}.
\ee
As above, this implies a
high gas density and thus a low Jeans mass,
\begin{equation}
M_* \approx M_J \propto T^{3/2} \rho^{-1/2}.
\ee
This results in the mass of the stars being directly related to their separation,
\be
R_{\rm sep} \propto \frac{M_*}{T},
\end{equation}
such that close systems have very low masses.  For example, if the typical 30 AU binary has solar mass
components, then a $1/3$ AU binary should have components of $0.01$ \solmas.
Forming close binary stars in situ is therefore difficult as
it requires subsequent accretion to reach stellar masses (Bonnell \& Bate 1994).
An alternative is that the components form at greater separation and then
are brought together.
Recent simulations of star formation in a cluster environment have shown that close binaries can result
from the induced evolution of wider systems (Bate, Bonnell \& Bromm 2003a). The binaries
evolve due to gas accretion and dynamical interactions with other stars. Can the same
processes explain high-mass close binary systems?

\section{The Formation of Stellar Clusters}	

One inescapable feature of massive stars is that they form in rich stellar clusters (Clarke \etal~2000; Lada \& Lada 2003. In fact,
with the exception or runaway stars, presumably ejected from clusters or binary systems,
there is little evidence for young massive stars not to be in the centre of dense stellar systems.
Furthermore, the central O stars of young open clusters are generally in close binary systems
with comparable mass companions (Mermilliod 2000).
It would therefore seem sensible to study the formation of massive stars in the context of the
formation of a stellar cluster. 

The fragmentation of a molecular cloud to form a stellar cluster has been the subject
of several studies (Boss 1996; Klessen et al. 1998; Klessen \& Burkert 2000; Bate, Bonnell \& Bromm 2003b; Bonnell, Bate \& Vine 2003).  In recent work, the fragmentation is
due to turbulently generated structure in molecular clouds. For example, in Bonnell et al. (2003),
the fragmentation of a 1000 \solmas\ cloud occurs as the turbulence generates filamentary
structure (see Figure~1). The filaments fragment to form individual stars which fall into local potential
wells forming small clusters. These clusters grow by accreting further stars and gas
and eventually  merge to form a large stellar cluster containing some $\approx 400$ stars. This
simulation was also noteworthy as it was the first that was sufficiently large to populate
a full IMF from low-mass to high-mass stars.

\begin{figure}[!ht]
\plotone{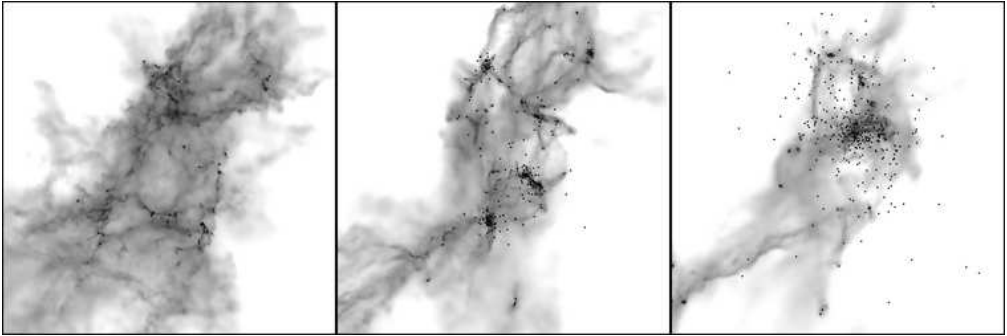}
\caption{The fragmentation of a 1000 \solmas\ molecular cloud and the formation of a stellar
cluster containing 419 stars (Bonnell \etal 2003).}
\end{figure}

Competitive accretion in stellar clusters occurs as individual stars compete gravitationally
for the reservoir of gas. As fragmentation is highly inefficient (eg Bate et al 2003b; Bonnell et al 2003),
there remains a large mass of gas that can dominate the potential. Accretion of this
gas can then determine the final stellar masses. Numerical studies have shown that
competitive accretion results in a large range of masses, with stars in the centre
of the cluster, the deepest part of the potential, accreting more gas due to their location
(Bonnell \etal~1997; Bonnell \etal~2001a). Furthermore, competitive accretion can
explain the initial mass function as it predicts a two power-law IMF.  Low-mass
stars accrete the majority of their mass in a gas dominated regime where tidal
forces limit their eventual mass, resulting in a shallow IMF (Bonnell \etal~2001b).  Higher-mass stars accrete their mass in the stellar dominated cores of the clusters with accretion rates determined by a Bondi-Hoyle process, resulting in  a steeper IMF.
 

Of added importance here is that accretion also increases the local stellar density. Infalling
mass that is accreted by individual stars increase the stars' binding energy forcing the cluster to contract
(Bonnell \etal 1998; Bonnell \& Bate 2002). In the study of Bonnell \& Bate (2002), gas accretion onto
a cluster of 1000 stars increases the core stellar density by a factor of over $10^5$. Thus
massive stars are expected to form in dense environments (See Figure~ 2a). This is
equivalent to a decrease in stellar separations by a factor of 50.  The formation of close massive binaries can occur in an analogous fashion (see below).

\subsection{Accretion and Massive Stars}   

The numerical simulations discussed above provides a framework in which to understand
the formation of massive stars. In this scenario, the massive stars form due to competitive 
accretion onto the core of the cluster in which the massive star is forming. The Lagrangian nature
of the SPH hydro code used for the simulations allows for the decomposition of the mass accretion
history of each star. In this way we have been able to analyse where the mass of the massive
stars comes from and thus the potential for accretion to form close binaries amongst the massive stars.

We found that the vast majority of the mass which comprises the massive stars comes
from large distances and is accreted onto the star after a stellar cluster has formed (see Figure~2b). The initial
fragment mass which forms the star is of low mass, typical to the mean stellar mass. The infalling
gas then has to pass through the cluster to be accreted by the central massive star. We also found
that the infalling gas is accompanied by newly formed stars such that the formation of a massive star
is a necessary byproduct of the formation of a stellar cluster. It should be noted that these simulations
neglect the effect of radiation pressure from the massive stars, or equivalently assume that
accretion through a disc (Yorke \& Sonnhalter 2002) occurs.

\begin{figure}[!ht]
\plottwo{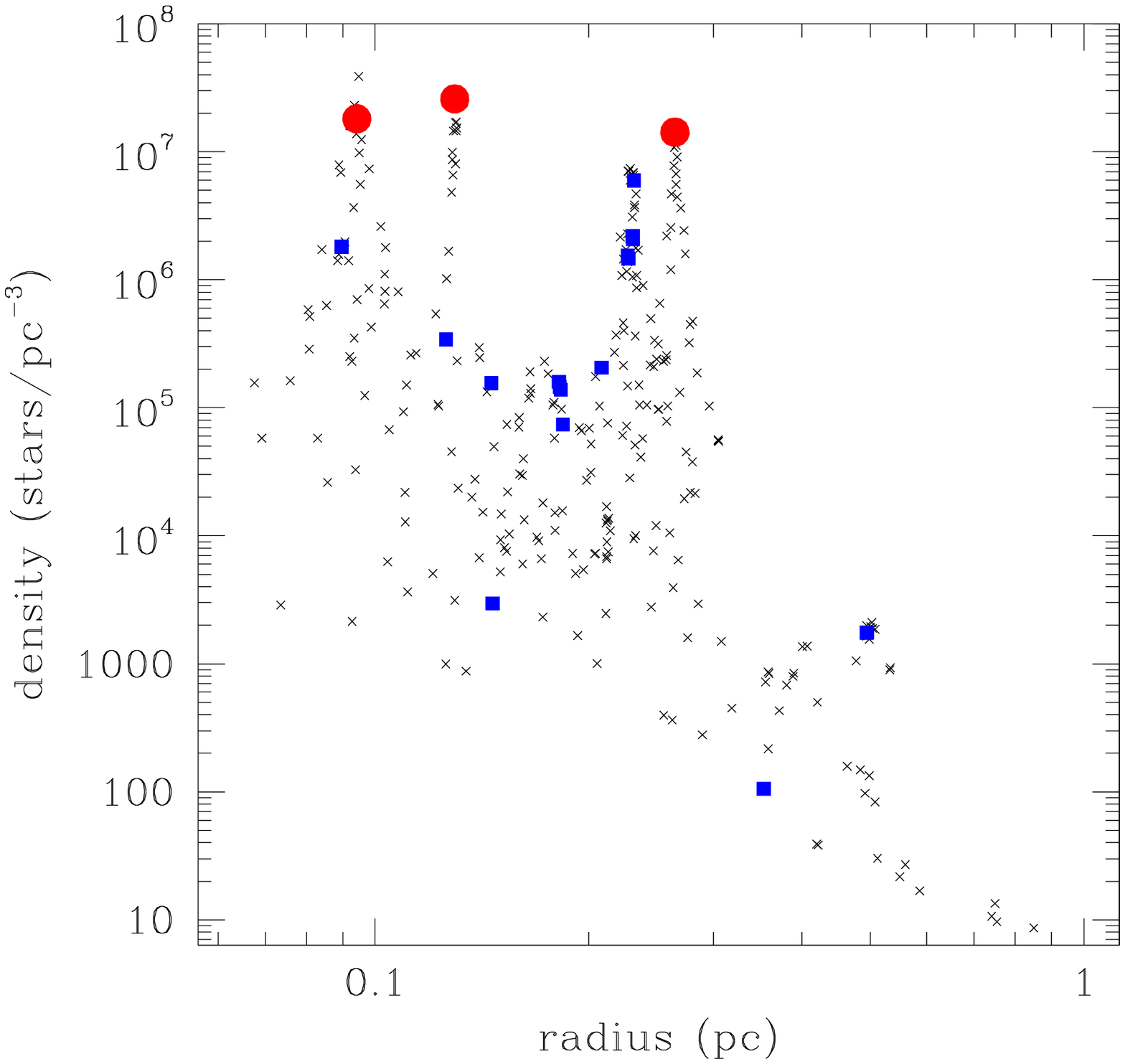}{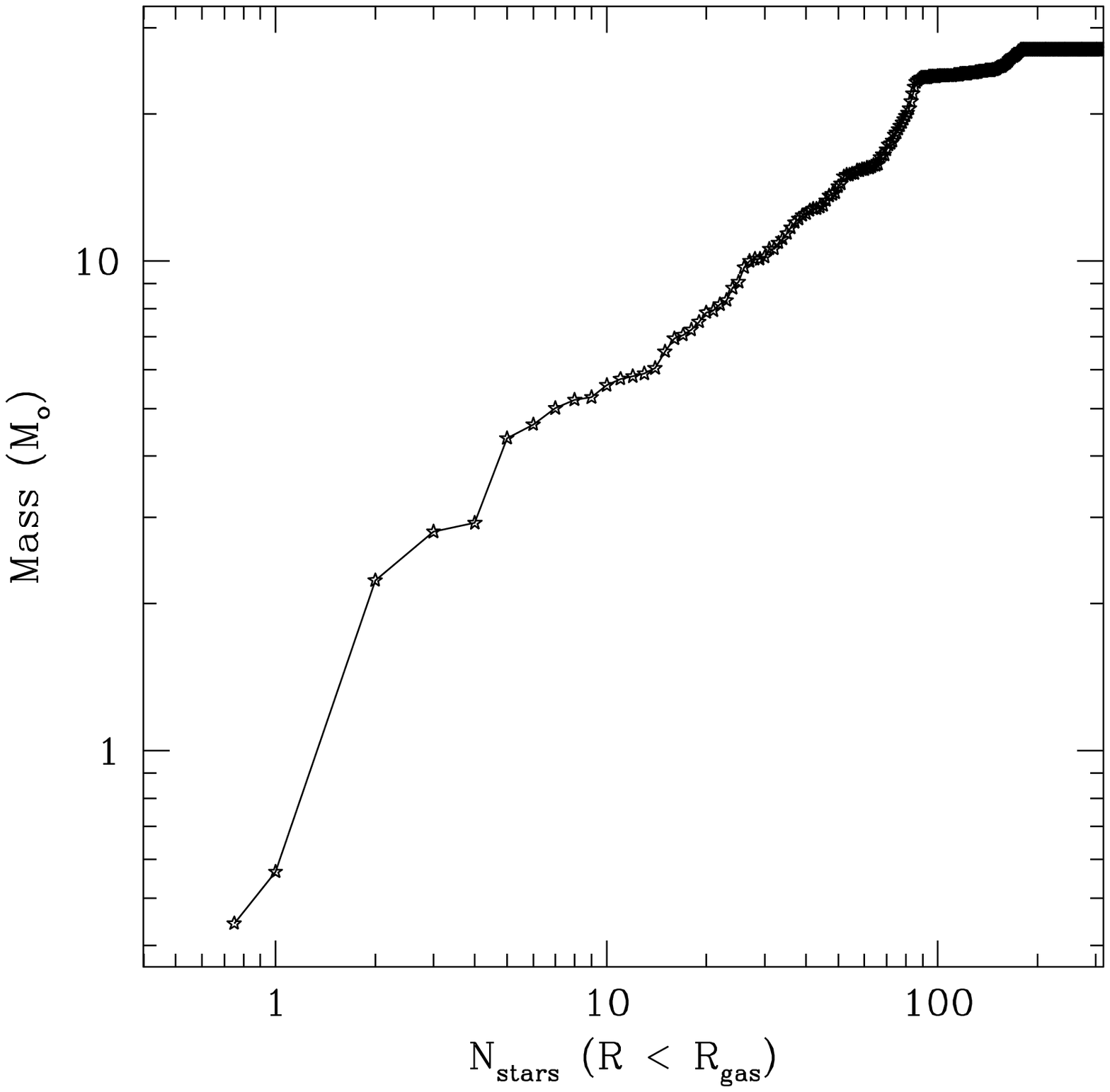}
\caption{ The stellar density is plotted against radius in the 400 star cluster in the left panel. Stars
with $m > 10$ \solmas\  are plotted as solid cricles while those more massive than 2 \solmas\  are plotted as squares. The right panel shows that the majority of the final stellar mass comes from outside the central binary and even outside the forming cluster of stars (Bonnell \etal~2004).}
\end{figure}

\section{Binary Formation and Evolution}   

The numerical simulations detailed above form a significant number of binary systems amongst the
massive stars. The
binary systems generally form through three-body capture in the cores of the clusters. During the
earlier stages of the cluster formation, the small number of stars contained in each subcluster allows for significant interactions and a relatively low velocity dispersion such that three-body capture is 
common and few of the eventual higher-mass stars are not in binary or multiple systems.

\subsection{Accretion and Binary Evolution}   

Accretion onto binary systems has the potential of forming close systems out of
wider systems at the same time as forming higher-mass components. In order to see this, let
us consider the angular momentum of a binary system,
\be
J \propto M^{3/2} R^{1/2}.
\ee
If the accreted material has zero net angular momentum as is expected if it infalls
spherically, then this implies that the binary separation should be a strong function of the mass,
\be 
R \propto M^{-3}.
\ee
If instead, the accreted material has constant specific angular momentum, the same as the
initial binary, then $J \propto M$ and
\be
R \propto M^{-1}.
\ee

\begin{figure}[!ht]
\plottwo{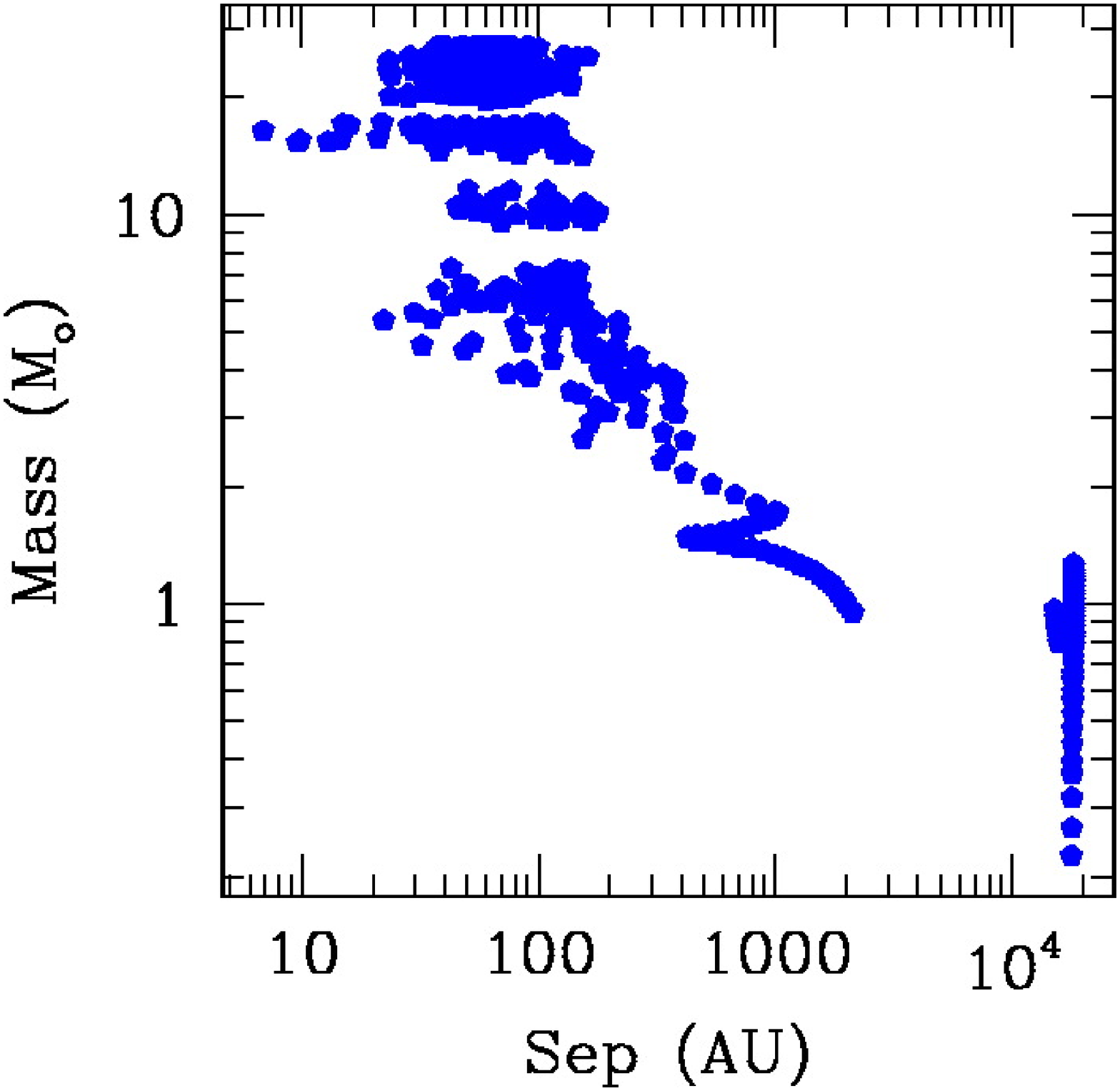}{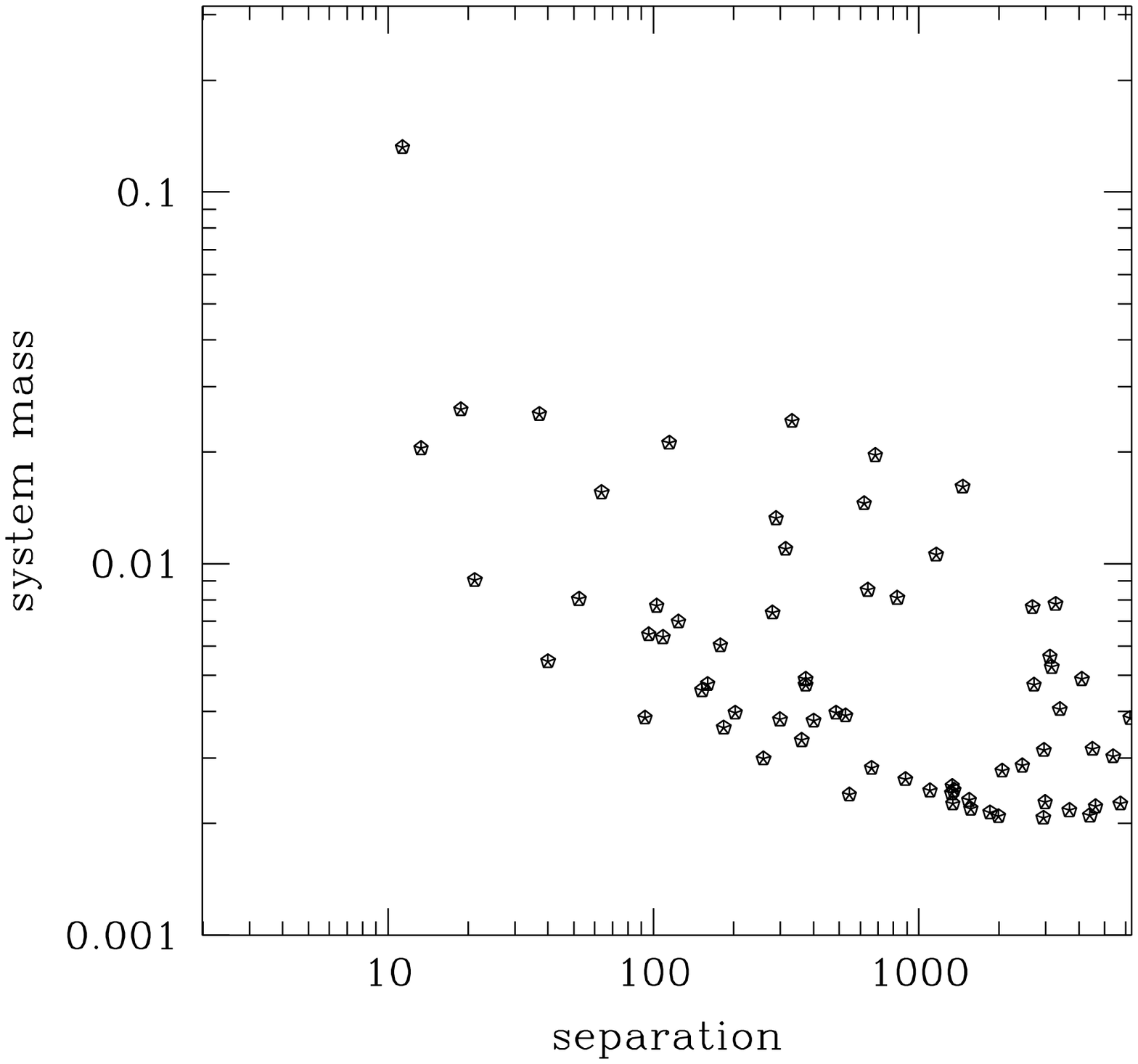}
\caption{ The evolution of a massive binary's mass versus its separation due to the
combined effect of accretion and stellar interactions. The gravitational softening 
is set at 180 AU, such that all separations below this value are upper limits. The panel on the right
plots the final distribution of binary masses (in units of 1000 \solmas, as a function of their 
separation in units of AU (Bonnell \& Bate 2002). We see that the separation decreases
for the more massive systems which have undergone the most accretion.}
\end{figure}

Thus we can see that a large decrease in the orbital separation can occur when
significant mass accretion forms a high-mass binary system. Unfortunately, in most of the
simulations reported above, the gravitational potentials were smoothed in order
to minimise computational expense. Still, we can see that the accretion is having a significant
effect on the binary's evolution. Figure~3a  plots the evolution of a binary's separation versus
its mass due to gas accretion. We see that the separation decreases dramatically as the
mass increases. Even though the separation is limited by the gravitational
smoothing of 180 AU, we can estimate that the separation decreases with system mass
as 
\be
R\propto M^{-2}
\ee 
such that an increase in mass from typical stellar masses of 1 \solmas to high-masses
of $50$ \solmas would decrease a binary's separation from 1000 AU to 0.4 AU. Thus, forming
close systems from accretion is feasible (Figure 3b),

\subsection{Dynamics and Binary Evolution}   

\begin{figure}[!ht]
\plotfiddle{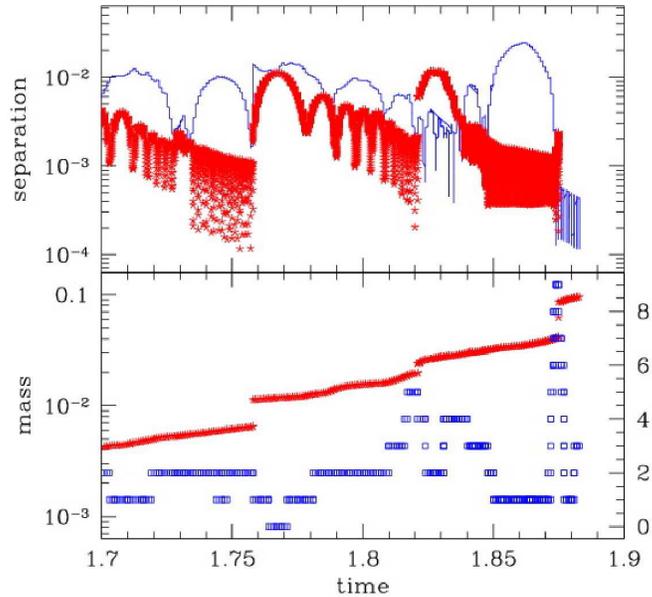}{8cm}{0}{45}{45}{-120}{-10}
\caption{ The dynamical hardening of a massive binary due to accretion and stellar interactions is
shown in the top panel. The binary's separation decreases steadily due to accretion but
can suddenly decrease due to the passage of a third star. Occasionally this hardening forces
the binary to merge, forming a higher-mass star with the perturbing star now becoming the companion. 
The mass (upper curve) of the primary is plotted in the bottom panel as well as the number of stars (Lower curve) in its vicinity. Jumps in the mass indicate stellar mergers (Bonnell \& Bate 2002).}
\end{figure}

A secondary effect of the accretion in clusters is to increase the stellar density.
This occurs in an analogous manner to the evolution of a binary system undergoing
accretion depending on the amount of momentum imparted with the mass to the
accreting stars (Bonnell \etal 1998). As discussed above, simulations of accreting
clusters show increases in stellar densities of order $10^5$. This has two important
effects. Firstly, it increases the probability of stars passing close to a  binary
and thus hardening it by removing some of the binary's orbital angular momentum.
This effect has been noted to be relevant for the formation of close low-mass binaries (Bate \etal 2003a).
In terms of forming massive close binaries, the simulations show that the massive stars
always form in the centre of the clusters where the potential well is deepest, It is also here
that the stellar density is highest  (Figure~ 2a) and thus binary hardening through stellar interactions
is most likely even more important in the case of high-mass systems.

\begin{figure}[!ht]
\plotone{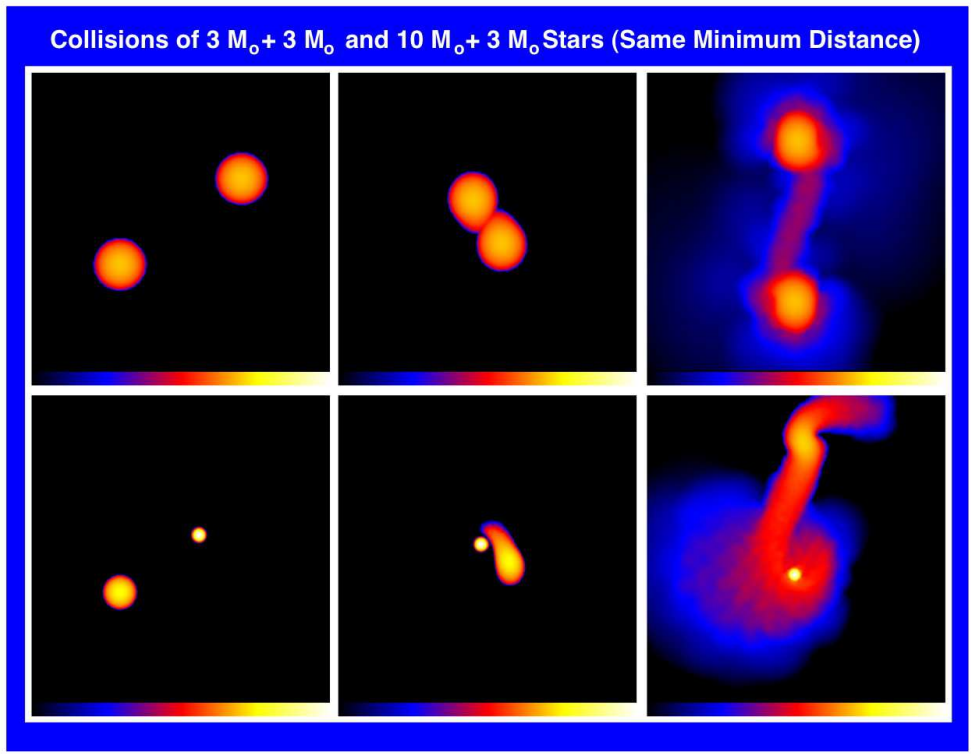}
\caption{ the interactions between two 3 \solmas stars (top) which results
in a tidal capture while the interaction between a 10 and a 3 \solmas stars 
results in the tidal disruption of the 3 \solmas\ star (Davies \etal~2004).}
\end{figure}

Binary hardening is limited by the size of the stars. In extreme cricumstances, the hardening
will force the binary to merge to form a higher mass system. This was seen to occur
in the simulations of accretion onto a large stellar cluster (Bonnell \& Bate 2002). 
Figure~4 plots the evolution of a massive binary system due to the presence of additional stars
in the cluster core. The binary's separation evolves due to ongoing accretion and due
to the perturbation of other stars in the core. Accretion leads to a steady smooth
decline in the binary separation while close interactions lead to a rapid decrease in the
separation and sometimes to direct mergers. At this point, the third star, generally already
bound due to the accretion and stellar dynamics in the cluster core, becomes the 
binary companion to a more massive primary.

Two other binary formation processes can occur in dense stellar clusters. One
is the aforementioned tidal capture when unbound stars pass within  a few stellar radii
of each other (Fabian, Pringle \& Rees 1975; Bonnell \etal 1998). The second process
is disc capture of passing stars (Clarke \& Pringle 1991; Hall, Clarke \& Pringle 1996).
These processes require the presence of a circumstellar disc which can be due to either
direct infall, or due to the tidal disruption of a low-mass star (Davies \etal 2004). In Davies
\etal~(2004), we investigate the stellar collisions expected in a merger scenario and find that
while two high-mass stars are capable of tidal capture, a high-mass interacting with a
pre-main sequence low mass star results in a tidal disruption of the low mass star. This
material will form an inner disc which can then act to aid in the capture
of any successive stellar interactions (Figure~5).

\section{Conclusions}

The formation of close massive binary stars can be understood in the context of
the formation of a stellar cluster and the subsequent competitive accretion which
produces the initial mass function. Massive stars form in the cores of the cluster where
three-body capture forms binary systems. These binaries evolve from wide
low-mass systems to close high-mass systems due to the effects of gas accretion and stellar
interactions and binary hardening. Binary mergers due to such hardening may be
important in overcoming the effects of radiation pressure. The expected results of
such processes are close systems with comparable mass companions. The mass ratios
should be fairly high due to mass accretion  favouring equal mass ratios and due
to any exchanges from stellar interactions.




\begin{thebibliography}{}

\bibitem[]{} Beech M., Mitalas R., 1994, ApJS, 95, 517

\bibitem[]{} Bate, M. R., Bonnell, I. A., Bromm, V. 2003a, MNRAS, 336, 705 

\bibitem[]{} Bate, M. R., Bonnell, I. A., Bromm, V. 2003b, MNRAS, 339, 577

\bibitem[]{} Bonnell, I. A., Bate, M. R.,  1994, MNRAS,  271, 999

\bibitem[]{} Bonnell, I. A., Bate, M. R.,  2002, MNRAS,  336, 659

\bibitem[]{} Bonnell, I. A., Bate, M. R.,  Clarke, C. J., \& Pringle, J. E., 1997, MNRAS, 285, 201

\bibitem[]{} Bonnell, I. A., Bate, M. R.,  Clarke, C. J., \& Pringle, J. E., 2001a, MNRAS, 323, 785

\bibitem[]{} Bonnell, I. A., Bate, M. R., \& Vine, S. G. 2003, MNRAS, 343,413

\bibitem[]{} Bonnell, I. A., Bate, M. R., \& Zinnecker, H., 1998, MNRAS, 298, 93

\bibitem[]{} Bonnell, I. A., Clarke, C. J., Bate, M. R.,  \& Pringle, J. E., 2001b, MNRAS, 324, 573

\bibitem[]{} Bonnell, I. A., Vine, S. G., \& Bate, M. R. 2004, MNRAS, 349, 735

\bibitem[]{} Boss, A.P., 1996, ApJ, 468, 231

\bibitem[]{} Clarke, C. J., Bonnell, I. A., Hillenbrand, L. A., 2000, PPIV,  Manni
ngs, Boss \& Russell eds, p 151

\bibitem[]{} Clarke, C.J., Pringle, J.E., 1991, MNRAS, 249, 584

\bibitem[]{} Davies, M.B., Bate, M.R., Bonnell, I.A., 2004, in preparation

\bibitem[]{} Fabian, A.C., Pringle, J.E., Rees, M.J., 1975, MNRAS, 

\bibitem[]{} Hall, S., Clarke, C.J., Pringle, J.E., 1991, MNRAS, 278, 303

\bibitem[]{} Klessen R.~S., Burkert A., 2000, ApJS, 
128, 287 

\bibitem[]{} Klessen R.~S., Burkert A., Bate M.~R., 
1998, ApJ, 501, L205 

\bibitem[]{} Lada C. J., Lada, E. 2003, ARA\&A, 41, 57

\bibitem[]{} McKee, C.F.,  Tan, J.C., 2003, ApJ, 585, 850

\bibitem[]{} Mermilliod J.C., 2001, in IAU 200, eds H. Zinnecker and R. Mathieu
\bibitem[]{} Wolfire M.G., Cassinelli J.P., 1987, ApJ, 319, 850
\bibitem[]{} Yorke H., Sonnhalter, C., 2002, ApJ, 569, 846
\end{thebibliography}
\end{document}